\documentclass[a4paper,11pt,aps,prd,reprint]{revtex4-1}

\usepackage{amsmath,amssymb,amsfonts} 

\usepackage[usenames,dvipsnames,svgnames]{xcolor}
\usepackage[colorlinks,
	linkcolor=red,
	citecolor=blue,
	urlcolor=red]{hyperref}
\usepackage{graphicx}

\usepackage{cleveref}

\newcommand{\re}{\mathrm{Re}\,}
\newcommand{\im}{\mathrm{Im}\,}
\renewcommand{\vec}[1]{\mathbf{{#1}}}

\renewcommand{\t}[1]{\mathrm{#1}}
\renewcommand{\d}{\mathrm{d}}
\newcommand{\abs}[1]{\left \vert {#1} \right \vert}

\begin{document}

\title{Relativistic coupling of phase and amplitude noise in optical interferometry}
\author{Vivishek Sudhir, Peter Fritschel, Nergis Mavalvala}
\affiliation{LIGO Laboratory, Massachusetts Institute of Technology, Cambridge, USA}
\date{\today}

\begin{abstract}
Extraneous motion of optical elements in an interferometer lead to excess noise.
Typically, fluctuations in the effective path length lead to phase
noise, while beam-pointing leads to apparent amplitude noise.
For a transmissive optic moving along the optical axis, neither effect should exist.
However, relativity of motion suggests that even in this case, small corrections of order $v/c$ ($v$ the velocity
of the optic), give rise to phase and amplitude noise on the light. Here we calculate the 
effect of this relativistic mechanism of noise coupling, and discuss when such an effect would limit 
the sensitivity of optical interferometers.
\end{abstract}

\maketitle

\section{Introduction}

More than a century ago, Fizeau had observed that the velocity of light in a moving dielectric medium
was reduced by a factor proportional to the velocity of the medium \cite{Fiz59}.
Although wrongly interpreted as proof of a luminiferous aether (see for example \cite{MichMor86}),
it was soon realized that the effect was due to the relativity of motion of the medium on
the phase of the light \cite{Lau07,Pauli21}.
Instead of being in uniform motion, if the medium were to be subjected to
random motion -- either of technical or fundamental origin -- the resulting random velocity of
light would cause a random phase shift on it.
The consequence of this is that any optical interferometer with partially transmissive optical elements would
be susceptible to this relativistic source of phase noise \cite{HarFrit07}.

In addition to noise on the longitudinal mode of the propagating optical field that leads to phase noise,
motion of the optic can also distort the transverse (spatial) mode by relativistic aberration \cite{Pauli21,Mor94}.
For the optic in random motion, the aberration is also random. When interfered with a reference beam at
a photodetector, the fluctuations in the transverse mode lead to fluctuations in the beam overlap, leading to
apparent amplitude noise.

These two relativistic effects, taken together form a source of apparent phase and amplitude noise in optical
interferometers with transmissive optics that is of fundamental origin. This work thus complements previous
analysis of fundamental sources of optical noise, of thermodynamic \cite{Saul90,Glenn89,Brag99,Brag00,Harry12}
or quantum mechanical origin \cite{Caves81}.
As we shall show, the relativistic sources of noise studied here are far from challenging the capability of
today's most sensitive optical interferometer, Advanced LIGO \cite{ligo16}. However, the physics studied here
may impact the design of future interferometers.

\section{Model and theory}

Conventionally, the amplitude and phase of a light beam of interest -- called the ``signal'' --
say propagating along the $z$-axis, is defined via the decomposition of its
electric field $E_s(\vec{r},t)$ into the form \cite{Sieg86} (for a fixed, but arbitrary polarization),
\begin{equation}\label{eq:E}
  E_s(\vec{r},t) = u_s(\vec{r}) \exp \left[ i\phi_s(z,t) \right],
\end{equation}
where, $\phi_s(z,t)=kz-\omega t$, is the longitudinal mode which gives the phase of the field, and $u_s(\vec{r})$ the
transverse mode whose magnitude gives the amplitude.
Fluctuations in the former (latter) are then said to constitute phase (amplitude) noise.
This identification however fails in the case considered here where relativity of motion of a dielectric 
optic (with finite optical thickness) 
is the underlying cause of fluctuations in the longitudinal and transverse modes. The reason is that the above
decomposition of the field is not Lorentz invariant, making the
identification of the amplitude and phase frame-dependent. It is therefore necessary to specify
the measurement that the beam is subjected to, and thence to operationally define what we mean by the
amplitude and phase of the beam, in terms of the tangible output of that measurement.

\begin{figure}[h!]
  \centering
  \includegraphics[width=0.8\columnwidth]{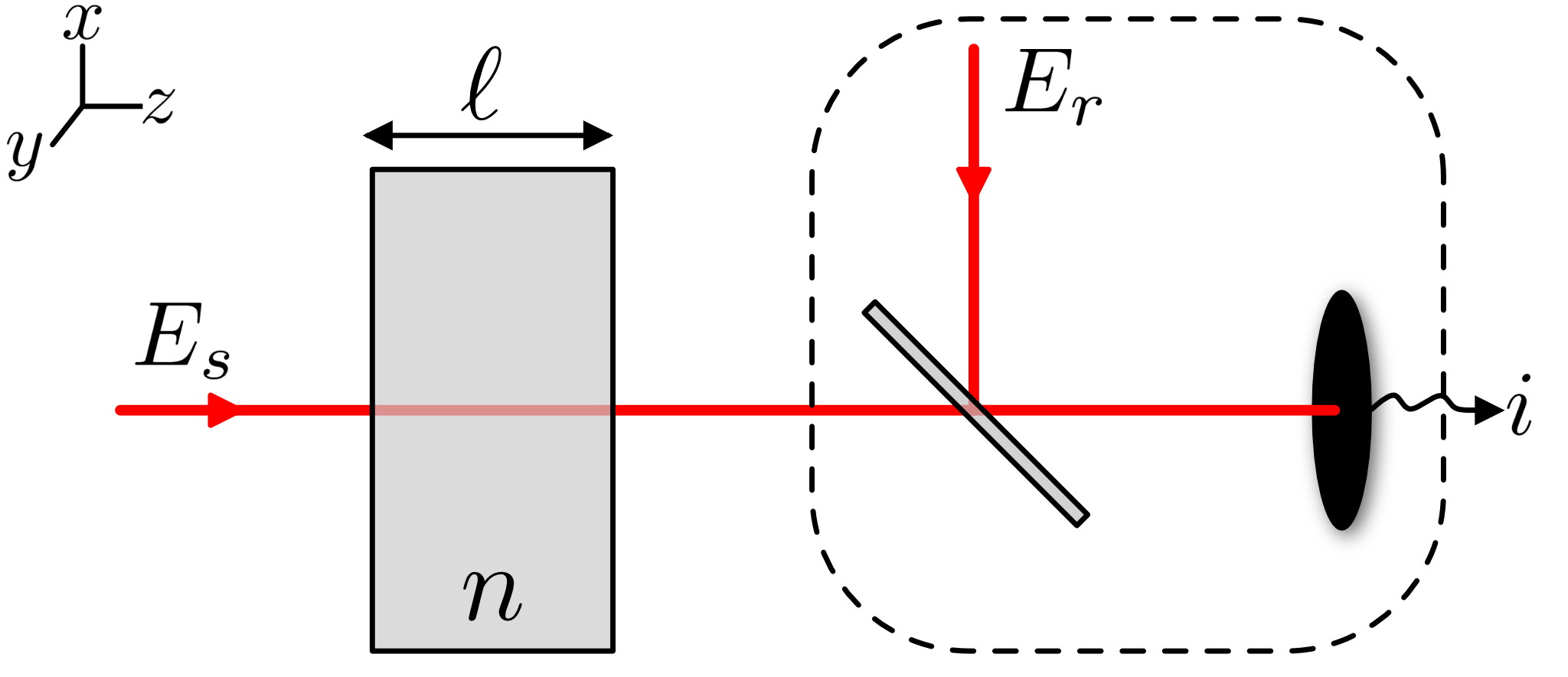}
  \caption{Sketch of interference configuration: signal field $E_s$ transmits through an optic of thickness $\ell$,
    and refractive index $n$, before interfering with a reference field $E_r$. Arrows denote direction of propagation.
  }
  \label{fig:scheme}
\end{figure}

Here we adopt the model depicted in \cref{fig:scheme}. The electric
field $E_s$ of the signal beam is transmitted through a movable optical element of thickness $\ell$ and
refractive index $n$ (in its rest frame) -- this is the element whose motion along $z$ leads to noise on the
signal. The output is mixed with a reference beam $E_r$ at a balanced beam-splitter, whose output is directed onto a
photo-detector -- this is the measurement apparatus, which is assumed motionless (and in fact defines what we mean
by the rest frame). We assume that the optic is suitably anti-reflection coated so as to ensure that a cavity is 
not formed across its thickness.

Fluctuations in the signal field are imprinted onto fluctuations of the photocurrent, viz.,
\begin{equation}\label{eq:di}
  \delta I(t) = \frac{\alpha}{2} \int_A \left( \bar{E}_\t{lo} \delta E_s^*
    + \bar{E}_\t{lo}^* \delta E_s \right)\, \d \vec{x},
\end{equation}
where $\alpha$ is the responsivity of the detector, $A$ is its active area, $\vec x = (x,y)$ are the coordinates
in the transverse plane, and $\bar{E}_\t{lo} = \bar{E}_r + \bar{E_s}$ is the mean local oscillator field that
is the sum of the mean reference and signal fields. Here the integrands are evaluated on the detector surface;
we implicitly assume this throughout.
We will henceforth also assume that $\bar{E}_r \gg \bar{E}_s$,
allowing us to neglect fluctuations in the reference, and to set $\bar{E}_\t{lo} \approx \bar{E}_r$.
Referring field fluctuations to fluctuations in its transverse and longitudinal modes,
\begin{equation}\label{eq:di1}
\begin{split}
  \delta I_{\bar{\phi}}(t) &\approx \alpha \int_A \abs{ \bar{u}_r \bar{u}_s }
  \left[ \left( \re \frac{\delta u_s}{\bar{u}_s} \right) \cos \bar{\phi} \right. \\
  &\qquad\qquad\qquad \left. + \left( \delta \phi_s + \im \frac{\delta u_s}{\bar{u}_s} \right) \sin \bar{\phi}
  \right] \d \vec{x},
\end{split}
\end{equation}
where, $\bar{u}_{r,s}$ are the mean reference and signal transverse mode functions,
and $\bar{\phi} \equiv \bar{\phi}_r -\bar{\phi}_s$ is the mean difference between their phases.

Amplitude (phase) fluctuations can now be operationally defined as those that lead to photocurrent fluctuations
arising when the mean reference field is in-phase (quadrature-phase) with the mean signal, i.e. $\bar{\phi} = 0$
($\bar{\phi}=\pi/2$). For example, the apparent relative power fluctuation in the signal $\delta \Pi_s$, can be
defined in terms of the photocurrent as,
\begin{equation}\label{eq:dPbyP}
  \delta \Pi_s(t) \equiv \frac{\delta I_0(t)}{\bar{I}}
  = \frac{2\int_A \abs{ \bar{u}_r \bar{u}_s }  (\re \delta u_s/\bar{u}_s) \d \vec{x}}
  {\int_A ( \abs{ \bar{u}_r}^2 + \abs{ \bar{u}_s}^2 ) \d \vec{x}},
\end{equation}
where, $\bar{I}=(\alpha/2)\int_A (\abs{\bar{u}_r}^2 + \abs{\bar{u}_s}^2)\, \d \vec{x}$, is the mean photocurrent.
Similarly, apparent phase fluctuations may be defined as,
\begin{equation}\label{eq:dPhi}
  \delta \Phi_s(t) \equiv \frac{\delta I_{\pi/2}(t)}{\bar{I}}
  = \frac{2\int_A \abs{ \bar{u}_r \bar{u}_s }( \delta \phi_s + \im \delta u_s/\bar{u}_s)\d \vec{x}}
  {\int_A ( \abs{ \bar{u}_r}^2 + \abs{ \bar{u}_s}^2 ) \d \vec{x}}.
\end{equation}
We thus have an operational framework for partitioning photocurrent fluctuations into apparent amplitude and
phase fluctuations.

In \Cref{sec:long} and \Cref{sec:trans}, we calculate the fluctuations in the longitudinal and transverse
mode, i.e. $\delta \phi_s$ and $\delta u_s$, due to relativistic Doppler shift and aberration respectively.
\Cref{sec:PhaseAmp} expresses these results in terms of amplitude and phase fluctuations using the operational
definitions given in \cref{eq:dPbyP,eq:dPhi}.

\subsection{Longitudinal mode fluctuations}\label{sec:long}

Consider the scenario where the signal field given in \cref{eq:E} propagates through a medium which is itself
moving with velocity $v$.
In its rest frame, denote by $\ell^\mu = (0,0,\ell,c\tau)$ its 4-thickness 
(where $\tau=\ell/(c/n)$ is the traversal time)
and by $k^\mu = (0,0,k,\omega/c)$ its 4-wave-vector (where $\omega=(c/n)k$ the frequency).
Similarly let $\ell'^\mu$ and $k'^\mu$ be the same quantities in the lab frame.
Lorentz invariance implies that,
\begin{equation}
  k'^\mu \ell'_\mu = k^\mu \ell_\mu.
\end{equation}
The 4-thickness of the medium in the two frames are related by the Lorentz transformations,
$\ell'^\mu = \Lambda^\mu_\nu \ell^\nu$, given by,
\begin{equation}
  \Lambda^3_3 =\Lambda^4_4 = \gamma,\, \Lambda^3_4 = \Lambda^4_3 = -u\gamma, 
\end{equation}
with other elements zero, and $\gamma = (1-u^2)^{-1/2}$, $u=v/c$.
Solving for the 4-wave-vector in the medium, we get,
\begin{equation}
  k' = \gamma (1-u/n)k, \quad \omega' = \gamma (1-nu)\omega.
\end{equation}
These equations simply express the relativistic Doppler effect in the medium,
which leads to the phase shift,
\begin{equation}
\begin{split}
  \delta \phi_s = (k'-k) \ell - (\omega' -\omega) \tau
  = \frac{\omega \ell}{c}(n^2 -1) u\gamma.
\end{split}
\end{equation}

If the velocity of the medium fluctuates about a mean value of zero, with fluctuations $\delta v$ small compared
to $c$, the phase shift takes the form,
\begin{equation}\label{eq:dphis}
  \delta \phi_s \approx
  \frac{\omega \ell}{c}(n^2 -1) \frac{\delta v}{c}.
\end{equation}
This results in an excess phase noise, quantified by the power spectral density at Fourier frequency $\Omega$,
\begin{equation}\label{eq:Sphis}
\begin{split}
  S_{\phi_s}[\Omega]
  & = \left(\frac{\omega}{c}\ell(n^2 -1)\frac{\Omega}{c} \right)^2 S_z[\Omega].
\end{split}
\end{equation}
where we have expressed the motion of the optic as a displacement spectral
density, $S_z[\Omega] = S_v[\Omega]/\Omega^2$.

\subsection{Transverse mode fluctuations} \label{sec:trans}

In addition to phase fluctuations that arise via fluctuations of the longitudinal mode, fluctuations of the transverse
spatial mode lead to apparent amplitude and phase fluctuations as per \cref{eq:dPhi}.

An ideal laser beam has a Gaussian transverse mode (in the paraxial approximation) \cite{KogLi66,Sieg86,LaxLou75}:
\begin{equation}\label{eq:u}
  u_s(\vec{x},z) = \sqrt{P_s} \frac{\exp\left[ik \abs{\vec{x}}^2/2Q(z) \right] }{Q(z)(i\pi Q(0)/k)^{1/2}},
\end{equation}
where, $Q(z)$ is the complex radius of curvature, given in terms of the real radius of curvature $R(z)$, and width
parameter $W(z)$, as $Q(z)^{-1} = R(z)^{-1} + i W(z)^{-1}$; $W(0)$ is related to the physical beam width at waist,
conventionally defined as the standard deviation of the Gaussian transverse mode and denoted $w(0)$,
as $W(0) = k w(0)^2$. Maxwell's equations in the paraxial approximation imply that the evolution of the transverse mode
along the beam axis $z$ is simply given by the relation, $Q(z) = z - iW(0)$, from which the evolution of the
radius of curvature and beam width can be derived.
The prefactor ensures the normalization convention that the transverse integral of the squared electric
field at the beam waist ($z=0$) gives the power $P_s$.
The transverse mode is thus fully described by the single complex function, $Q(z)$, and the optical power $P_s$.

Relativistic aberration causes $Q(z)$ to change as the beam traverses a moving optical element.
In order to calculate this change in $Q(z)$, we use the equivalence
between the wave optics picture of the transverse mode diverging from the beam waist, and the ray optics picture
of a pencil of rays emanating from the same point \cite{Somm54,KellStre71,Desch72}.
As is well known \cite{KogLi66,Sieg86}, this equivalence implies that the change in $Q$ at optical interfaces is
described by a conformal transformation whose elements are given by the ray transfer matrix -- this is the so-called
``ABCD formalism'' for beam propagation.

We now use the ray optic equivalence to calculate the ABCD matrix of the moving optic.
For the optic shown in \cref{fig:scheme}, 
the ray-transfer matrix $M$ is given by 4 factors, i.e. $M = M_1(n) M_2(\ell,u) M_3(n,u) M_4(n)$.
Here $M_4 = \t{diag}(1,1/n)$ describes the entrance interface, and $M_1 = \t{diag}(1,n)$ describes the exit interface.
The matrix $M_3$ describes the effect of relativistic aberration.
Rays that enter the moving medium non-parallel to the optical axis undergoe relativistic aberration, i.e.
an additional inclination of the ray in the lab frame; specifically, the inclination angle $\theta'$ is given by
\begin{equation}\label{eq:theta}
  \sin \theta' = \frac{(c/n) \sin \theta}{\gamma (cu+(c/n) \cos \theta)},
\end{equation}
where, $\theta$ is the angle of incidence.
This relation is the usual relativistic aberration equation \cite{Pauli21}, taking into account the optical density of
the medium via the velocity $c/n$.
For small angles, in keeping with the paraxial approximation, and for the velocity of the medium
small compared to $c$, this gives, $\theta' \approx \theta(1-n u)$. Thus,
\begin{equation}\label{eq:M3}
  M_3 (n,u) \approx \t{diag}(1,1-nu).
\end{equation}
The matrix $M_2$ describes the traversal of the beam in the optically dense medium; since the medium is moving,
its thickness in the lab frame is given by, $\ell/\gamma$, and so,
\begin{equation}\label{eq:M2}
  M_2 (\ell,u) = \begin{bmatrix} 1 & \ell/\gamma \\ 0 & 1 \end{bmatrix}
  \approx \begin{bmatrix} 1 & \ell \\ 0 & 1 \end{bmatrix};
\end{equation}
here, we have made an approximation to the first order in $u$.
Putting all this together, we get the transfer matrix for the moving medium,
\begin{equation}\label{eq:M}
  M \approx \begin{bmatrix} 1 & \frac{\ell}{n}(1-nu) \\ 0 & 1-nu \end{bmatrix}.
\end{equation}

The transfer matrix $M$ can now be applied to the input complex radius of curvature to determine the change in
the transverse mode profile. For simplicity, we take the beam waist to coincide with the entrance face of the optic,
so that the output transverse mode is characterized by,
\begin{equation}\label{eq:Qd}
\begin{split}
  Q(\ell) = \frac{M_{11} Q(0) + M_{12}}{M_{21} Q(0) + M_{22}}
   \approx Q(0)(1+nu) +\frac{\ell}{n}.
\end{split}
\end{equation}
For velocity fluctuating about zero, the transverse mode fluctuates by,
\begin{equation}\label{eq:dQ}
  \delta Q(\ell) \approx Q(0) n \frac{\delta v}{c},
\end{equation}
around the mean, $\bar{Q}(\ell) \equiv Q(0) + (\ell/n)$.
Physically, this corresponds to fluctuations in the real radius of curvature and width of the exiting beam,
implying a relative fluctuation in the transverse mode,
\begin{equation}\label{eq:du}
\begin{split}
  \frac{\delta u_s(\vec{x},\ell)}{\bar{u}_s(\vec{x},\ell)}
  &= -\left(1+i\frac{k\abs{\vec{x}}^2}{2\bar{Q}(\ell)}\right) \frac{\delta Q(\ell)}{\bar{Q}(\ell)} \\
  &\approx -\left(1+i\frac{k\abs{\vec{x}}^2}{2\bar{Q}(\ell)}\right)\left(\frac{n Q(0)}{\bar{Q}(\ell)}\right)
  \frac{\delta v}{c}.
\end{split}
\end{equation}

\subsection{Phase and amplitude noise in an interferometric measurement}\label{sec:PhaseAmp}

\Cref{eq:dphis,eq:du} describe the two facets of relativistic optical noise. Using these in the
operational definition of amplitude and phase described earlier, we may now derive expressions for the apparent
phase and amplitude noise in an interferometric measurement of the beam.

We assume that the transverse modes of the reference and signal beams are perfectly mode-matched at the detector,
and only differ in the optical power they carry, which we denote by $P_r$ and $P_s$ respectively.
With this assumption, and using the expressions for the mode fluctuations (\cref{eq:dphis,eq:du}) in the
expressions for the amplitude and phase fluctuations (\cref{eq:dPbyP,eq:dPhi}), we obtain,
\begin{equation}\label{eq:dPiPhi}
  \begin{split}
    S_{\Pi_s} &= \eta \left(n + \frac{2\epsilon^2}{\epsilon^2 +1} \right)^2 \frac{S_v}{c^2}\\
    S_{\Phi_s} &= \eta S_{\phi_s} + \eta \epsilon^2\left(n-\frac{\epsilon^2 -1}{\epsilon^2 +1} \right)^2
    \frac{S_v}{c^2},
  \end{split}
\end{equation}
where, $\eta \equiv (2\sqrt{P_r P_s}/(P_r + P_s))^2 \leq 1$ is the square of the contrast of the interference, and
$\epsilon \equiv \ell/W(0)$ is the thickness of the optic in units of the beam width parameter at
waist. Note that these expressions assume $\delta v \ll c$ (see below for caveats).

\section{Discussion}

In order to ascertain which of the contributions in \cref{eq:dPiPhi} is the largest, it is best to express
the dimensionless optic thickness $\epsilon$ in the form, $\epsilon = 2\pi(\ell/\lambda) (\lambda/w(0))^2$, where
$w(0) = \sqrt{W(0)/k}$ is the physical beam width at waist, and $\lambda=2\pi/k$ is the wavelength.
Diffraction constrains the beam-width to $w(0) \gtrsim \lambda$; further, for typical optic thicknesses,
$\ell \sim (1-10)\, \t{cm}$, and wavelength, $\lambda \sim 1\, \mu \t{m}$, we get that $\epsilon \ll 1$.
In this regime, it is clear from \cref{eq:dPiPhi} that the contribution of the transverse mode
fluctuations to the phase noise is negligible, i.e. $S_{\Phi_s} \approx \eta S_{\phi_s}$.
Furthermore, the magnitude of phase noise is larger than that of the amplitude noise,
i.e. $S_{\Phi_s} \gg S_{\Pi_s}$.

Relativistic phase noise, although dominating over its amplitude counterpart, is still small compared to
other noise sources in even the most precise interferometers of today.
Consider for example, the propagation of a beam of wavelength $1\, \mu\t{m}$ through
an optical element of refractive index $n=1.45$, thickness $d=6\, \t{cm}$, and vibrationally isolated using
a triple pendulum as the main beam-splitter in the Advanced LIGO detector \cite{aligo15_sei} for which the
residual displacement noise is,
$S_z^{1/2}[2\pi\cdot 10\, \t{Hz}] \approx 10^{-12} \t{m}/\sqrt{\t{Hz}}$. In this case, the relativistic phase
noise contribution is, $S_{\phi_s}^{1/2}[2\pi\cdot 10\, \t{Hz}] \approx 10^{-15}\, \t{rad}/\sqrt{\t{Hz}}$,
which is about 4 orders of magnitude smaller than the phase-noise-equivalent differential length fluctuation
requirement for Advanced LIGO at these frequencies \cite{ligo16}.

Another area of application where relativistic phase noise may seem to be relevant is the transfer of optical
frequency standards through fiber links \cite{ForYe07}. Calculations indicate that the various thermal sources of
phase noise do scale as the square of the length of the fiber \cite{ShelLev85,Wans92,Duan12},
similar to relativistic phase noise; nevertheless their absolute value at room temperature is about 6 orders of
magnitude larger.

In closing, we would like to highlight a few limitations of the theoretical treatment presented here.
Firstly, all expressions for the phase and amplitude noise power spectral densities rely on a linear approximation
in $v/c$; this means that all these expressions are only valid up to Fourier frequency $\Omega_\t{max}$, defined by,
$\int_0^{\Omega_\t{max}} S_v[\Omega]\, d\Omega/2\pi \ll c^2$. Beyond this frequency, the linear approximation in
$v/c$ breaks down. Secondly, one should also expect that for velocity fluctuations large enough, the resulting
deviation from non-inertial motion due to the large acceleration becomes important. Thirdly, the notion of a
rigid body is abhorrent to the principle of relativity \cite{Pauli21,Born09,Herg11},
thus the assumption of a rigid optic executing center-of-mass motion needs to be qualified.
In particular, such an approximation is valid for frequencies below the internal (elastic) resonances of the
optic. 
Fourthly, we have neglected the effect of chromatic dispersion in the analysis of phase noise. 
The effect of chromatic dispersion is to produce an additional refractive index fluctuation, 
$\delta n = (\partial n/\partial \lambda) \delta \lambda \approx (\partial n/\partial \lambda)\lambda (\delta v/c)$,
seeded by Doppler shift of the wavelength. In the case of fused silica, for which
$\partial n/\partial \lambda \approx -(0.01-0.03)/\mu \t{m}$ at $\lambda = 1\, \mu \t{m}$, 
chromatic dispersion thus gives a few percent excess contribution to the relativistic phase noise estimate.
Finally, we point out the amusing possibility of displacement of the optic induced by the light passing
through it -- fundamentally, either by radiation reaction on the dielectric \cite{DalCoh82},
or by gravitational stresses induced by the light itself \cite{Tol31,Scu79}.
These mechanisms may be seen as relativistic sources of back-action complementary to the transduction 
mechanism studied above.

\section{Conclusion}

Here, we have considered phase and amplitude noise arising from the motion of optical elements within the
context of special relativity. Conceptually, this completes the set of optical noises induced by a moving element
due to fundamental sources: thermodynamic fluctuations of the center-of-mass and of microscopic modes 
give rise to thermo-mechanical \cite{Saul90}, thermo-elastic \cite{Glenn89,Brag99}, or
thermo-refractive \cite{Glenn89,Brag00} noises;
quantum fluctuations of the radiation pressure can contaminate phase measurements \cite{Caves81}; we now know the
magnitude of noise added by effects arising due to relativity of motion.
In contemporary interferometers, the magnitude of these fundamental noises follows the same order as they are
written above.
Indeed, quantum fluctuations of radiation pressure have only been experimentally observed in the
last few years \cite{Purdy13,WilSud15}.
The relativistic source of noise is far from the limit posed for any contemporary optical interferometer;
however, ambitious future designs for interferometers targeting broadband sensitivity, even on the table-top,
could be affected by these relativistic effects if extraneous motion of optics is not suppressed.

\section*{Acknowledgements} VS thanks Rai Weiss for helpful discussions; and him and Chris Galland for a careful
reading of the manuscript. VS is supported by a research fellowship (no. $\t{P2ELP2}\_178231$) from the
Swiss National Science Foundation.
PF acknowledges the support of the National Science Foundation and the LIGO Laboratory, operating 
under cooperative agreement no. PHY-0757058.


\bibliographystyle{apsrev4-1}
\bibliography{refs_LO_phase}

\begin{thebibliography}{31}%
\makeatletter
\providecommand \@ifxundefined [1]{%
 \@ifx{#1\undefined}
}%
\providecommand \@ifnum [1]{%
 \ifnum #1\expandafter \@firstoftwo
 \else \expandafter \@secondoftwo
 \fi
}%
\providecommand \@ifx [1]{%
 \ifx #1\expandafter \@firstoftwo
 \else \expandafter \@secondoftwo
 \fi
}%
\providecommand \natexlab [1]{#1}%
\providecommand \enquote  [1]{``#1''}%
\providecommand \bibnamefont  [1]{#1}%
\providecommand \bibfnamefont [1]{#1}%
\providecommand \citenamefont [1]{#1}%
\providecommand \href@noop [0]{\@secondoftwo}%
\providecommand \href [0]{\begingroup \@sanitize@url \@href}%
\providecommand \@href[1]{\@@startlink{#1}\@@href}%
\providecommand \@@href[1]{\endgroup#1\@@endlink}%
\providecommand \@sanitize@url [0]{\catcode `\\12\catcode `\$12\catcode
  `\&12\catcode `\#12\catcode `\^12\catcode `\_12\catcode `\%12\relax}%
\providecommand \@@startlink[1]{}%
\providecommand \@@endlink[0]{}%
\providecommand \url  [0]{\begingroup\@sanitize@url \@url }%
\providecommand \@url [1]{\endgroup\@href {#1}{\urlprefix }}%
\providecommand \urlprefix  [0]{URL }%
\providecommand \Eprint [0]{\href }%
\providecommand \doibase [0]{http://dx.doi.org/}%
\providecommand \selectlanguage [0]{\@gobble}%
\providecommand \bibinfo  [0]{\@secondoftwo}%
\providecommand \bibfield  [0]{\@secondoftwo}%
\providecommand \translation [1]{[#1]}%
\providecommand \BibitemOpen [0]{}%
\providecommand \bibitemStop [0]{}%
\providecommand \bibitemNoStop [0]{.\EOS\space}%
\providecommand \EOS [0]{\spacefactor3000\relax}%
\providecommand \BibitemShut  [1]{\csname bibitem#1\endcsname}%
\let\auto@bib@innerbib\@empty
\bibitem [{\citenamefont {Fizeau}(1860)}]{Fiz59}%
  \BibitemOpen
  \bibfield  {author} {\bibinfo {author} {\bibfnamefont {H.}~\bibnamefont
  {Fizeau}},\ }\href {\doibase 10.1080/14786446008642856} {\bibfield  {journal}
  {\bibinfo  {journal} {Philosophical Magazine}\ }\textbf {\bibinfo {volume}
  {19}},\ \bibinfo {pages} {245} (\bibinfo {year} {1860})}\BibitemShut
  {NoStop}%
\bibitem [{\citenamefont {Michelson}\ and\ \citenamefont
  {Morley}(1886)}]{MichMor86}%
  \BibitemOpen
  \bibfield  {author} {\bibinfo {author} {\bibfnamefont {A.}~\bibnamefont
  {Michelson}}\ and\ \bibinfo {author} {\bibfnamefont {E.}~\bibnamefont
  {Morley}},\ }\href {\doibase 10.2475/ajs.s3-31.185.377} {\bibfield  {journal}
  {\bibinfo  {journal} {American Journal of Science}\ }\textbf {\bibinfo
  {volume} {185}},\ \bibinfo {pages} {377} (\bibinfo {year}
  {1886})}\BibitemShut {NoStop}%
\bibitem [{\citenamefont {von Laue}(1907)}]{Lau07}%
  \BibitemOpen
  \bibfield  {author} {\bibinfo {author} {\bibfnamefont {M.}~\bibnamefont {von
  Laue}},\ }\href {\doibase 10.1002/andp.19073281015} {\bibfield  {journal}
  {\bibinfo  {journal} {Annalen der Physik}\ }\textbf {\bibinfo {volume}
  {328}},\ \bibinfo {pages} {989} (\bibinfo {year} {1907})}\BibitemShut
  {NoStop}%
\bibitem [{\citenamefont {Pauli}(1958)}]{Pauli21}%
  \BibitemOpen
  \bibfield  {author} {\bibinfo {author} {\bibfnamefont {W.}~\bibnamefont
  {Pauli}},\ }\href@noop {} {\emph {\bibinfo {title} {Theory of Relativity}}}\
  (\bibinfo  {publisher} {Dover},\ \bibinfo {year} {1958})\BibitemShut
  {NoStop}%
\bibitem [{\citenamefont {Harry}\ and\ \citenamefont
  {Fritschel}(2007)}]{HarFrit07}%
  \BibitemOpen
  \bibfield  {author} {\bibinfo {author} {\bibfnamefont {G.}~\bibnamefont
  {Harry}}\ and\ \bibinfo {author} {\bibfnamefont {P.}~\bibnamefont
  {Fritschel}},\ }\href {https://dcc.ligo.org/LIGO-T010067} {\emph {\bibinfo
  {title} {Noise from the Fizeau effect}}},\ \bibinfo {type} {Tech. Rep.}\
  \bibinfo {number} {LIGO-T010067}\ (\bibinfo  {institution} {LIGO},\ \bibinfo
  {year} {2007})\BibitemShut {NoStop}%
\bibitem [{\citenamefont {Moreau}(1994)}]{Mor94}%
  \BibitemOpen
  \bibfield  {author} {\bibinfo {author} {\bibfnamefont {W.}~\bibnamefont
  {Moreau}},\ }\href {\doibase 10.1119/1.17543} {\bibfield  {journal} {\bibinfo
   {journal} {American Journal of Physics}\ }\textbf {\bibinfo {volume} {62}},\
  \bibinfo {pages} {426} (\bibinfo {year} {1994})}\BibitemShut {NoStop}%
\bibitem [{\citenamefont {Saulson}(1990)}]{Saul90}%
  \BibitemOpen
  \bibfield  {author} {\bibinfo {author} {\bibfnamefont {P.}~\bibnamefont
  {Saulson}},\ }\href {\doibase 10.1103/PhysRevD.42.2437} {\bibfield  {journal}
  {\bibinfo  {journal} {Physical Review D}\ }\textbf {\bibinfo {volume} {42}},\
  \bibinfo {pages} {2437} (\bibinfo {year} {1990})}\BibitemShut {NoStop}%
\bibitem [{\citenamefont {Glenn}(1989)}]{Glenn89}%
  \BibitemOpen
  \bibfield  {author} {\bibinfo {author} {\bibfnamefont {W.~H.}\ \bibnamefont
  {Glenn}},\ }\href {\doibase 10.1109/3.29251} {\bibfield  {journal} {\bibinfo
  {journal} {IEEE Journal of Quantum Electronics}\ }\textbf {\bibinfo {volume}
  {25}},\ \bibinfo {pages} {1218} (\bibinfo {year} {1989})}\BibitemShut
  {NoStop}%
\bibitem [{\citenamefont {Braginsky}\ \emph {et~al.}(1999)\citenamefont
  {Braginsky}, \citenamefont {Gorodetsky},\ and\ \citenamefont
  {Vyatchanin}}]{Brag99}%
  \BibitemOpen
  \bibfield  {author} {\bibinfo {author} {\bibfnamefont {V.~B.}\ \bibnamefont
  {Braginsky}}, \bibinfo {author} {\bibfnamefont {M.~L.}\ \bibnamefont
  {Gorodetsky}}, \ and\ \bibinfo {author} {\bibfnamefont {S.~P.}\ \bibnamefont
  {Vyatchanin}},\ }\href {\doibase 10.1016/S0375-9601(99)00785-9} {\bibfield
  {journal} {\bibinfo  {journal} {Physics Letters A}\ }\textbf {\bibinfo
  {volume} {264}},\ \bibinfo {pages} {1} (\bibinfo {year} {1999})}\BibitemShut
  {NoStop}%
\bibitem [{\citenamefont {Braginsky}\ \emph {et~al.}(2000)\citenamefont
  {Braginsky}, \citenamefont {Gorodetsky},\ and\ \citenamefont
  {Vyatchanin}}]{Brag00}%
  \BibitemOpen
  \bibfield  {author} {\bibinfo {author} {\bibfnamefont {V.~B.}\ \bibnamefont
  {Braginsky}}, \bibinfo {author} {\bibfnamefont {M.~L.}\ \bibnamefont
  {Gorodetsky}}, \ and\ \bibinfo {author} {\bibfnamefont {S.~P.}\ \bibnamefont
  {Vyatchanin}},\ }\href {\doibase 10.1016/S0375-9601(00)00389-3} {\bibfield
  {journal} {\bibinfo  {journal} {Physics Letters A}\ }\textbf {\bibinfo
  {volume} {271}},\ \bibinfo {pages} {303} (\bibinfo {year}
  {2000})}\BibitemShut {NoStop}%
\bibitem [{\citenamefont {Harry}\ \emph {et~al.}(2012)\citenamefont {Harry},
  \citenamefont {Bodiya},\ and\ \citenamefont {De~Salvo}}]{Harry12}%
  \BibitemOpen
  \bibinfo {editor} {\bibfnamefont {G.}~\bibnamefont {Harry}}, \bibinfo
  {editor} {\bibfnamefont {T.}~\bibnamefont {Bodiya}}, \ and\ \bibinfo {editor}
  {\bibfnamefont {R.}~\bibnamefont {De~Salvo}},\ eds.,\ \href@noop {} {\emph
  {\bibinfo {title} {Optical coatings and thermal noise in precision
  measurement}}}\ (\bibinfo  {publisher} {Cambridge University Press},\
  \bibinfo {year} {2012})\BibitemShut {NoStop}%
\bibitem [{\citenamefont {Caves}(1981)}]{Caves81}%
  \BibitemOpen
  \bibfield  {author} {\bibinfo {author} {\bibfnamefont {C.~M.}\ \bibnamefont
  {Caves}},\ }\href
  {https://journals.aps.org/prd/abstract/10.1103/PhysRevD.23.1693} {\bibfield
  {journal} {\bibinfo  {journal} {Physical Review D}\ }\textbf {\bibinfo
  {volume} {23}},\ \bibinfo {pages} {1693} (\bibinfo {year}
  {1981})}\BibitemShut {NoStop}%
\bibitem [{\citenamefont {{LIGO Scientific Collaboration and Virgo
  Collaboration}}(2016)}]{ligo16}%
  \BibitemOpen
  \bibfield  {author} {\bibinfo {author} {\bibnamefont {{LIGO Scientific
  Collaboration and Virgo Collaboration}}},\ }\href {\doibase
  10.1103/PhysRevLett.116.131103} {\bibfield  {journal} {\bibinfo  {journal}
  {Physical Review Letters}\ }\textbf {\bibinfo {volume} {116}},\ \bibinfo
  {pages} {131103} (\bibinfo {year} {2016})}\BibitemShut {NoStop}%
\bibitem [{\citenamefont {Siegman}(1986)}]{Sieg86}%
  \BibitemOpen
  \bibfield  {author} {\bibinfo {author} {\bibfnamefont {A.~E.}\ \bibnamefont
  {Siegman}},\ }\href@noop {} {\emph {\bibinfo {title} {Lasers}}}\ (\bibinfo
  {publisher} {University Science Books},\ \bibinfo {year} {1986})\BibitemShut
  {NoStop}%
\bibitem [{\citenamefont {Kogelnik}\ and\ \citenamefont {Li}(1966)}]{KogLi66}%
  \BibitemOpen
  \bibfield  {author} {\bibinfo {author} {\bibfnamefont {H.}~\bibnamefont
  {Kogelnik}}\ and\ \bibinfo {author} {\bibfnamefont {T.}~\bibnamefont {Li}},\
  }\href {\doibase 10.1364/AO.5.001550} {\bibfield  {journal} {\bibinfo
  {journal} {Applied Optics}\ }\textbf {\bibinfo {volume} {5}},\ \bibinfo
  {pages} {1518} (\bibinfo {year} {1966})}\BibitemShut {NoStop}%
\bibitem [{\citenamefont {Lax}\ \emph {et~al.}(1975)\citenamefont {Lax},
  \citenamefont {Louisell},\ and\ \citenamefont {McKnight}}]{LaxLou75}%
  \BibitemOpen
  \bibfield  {author} {\bibinfo {author} {\bibfnamefont {M.}~\bibnamefont
  {Lax}}, \bibinfo {author} {\bibfnamefont {W.~H.}\ \bibnamefont {Louisell}}, \
  and\ \bibinfo {author} {\bibfnamefont {W.~B.}\ \bibnamefont {McKnight}},\
  }\href {\doibase 10.1103/PhysRevA.11.1365} {\bibfield  {journal} {\bibinfo
  {journal} {Physical Review A}\ }\textbf {\bibinfo {volume} {11}},\ \bibinfo
  {pages} {1365} (\bibinfo {year} {1975})}\BibitemShut {NoStop}%
\bibitem [{\citenamefont {Sommerfeld}(1954)}]{Somm54}%
  \BibitemOpen
  \bibfield  {author} {\bibinfo {author} {\bibfnamefont {A.}~\bibnamefont
  {Sommerfeld}},\ }\href@noop {} {\emph {\bibinfo {title} {Optics}}}\ (\bibinfo
   {publisher} {Academic Press},\ \bibinfo {year} {1954})\BibitemShut {NoStop}%
\bibitem [{\citenamefont {Keller}\ and\ \citenamefont
  {Streifer}(1971)}]{KellStre71}%
  \BibitemOpen
  \bibfield  {author} {\bibinfo {author} {\bibfnamefont {J.~B.}\ \bibnamefont
  {Keller}}\ and\ \bibinfo {author} {\bibfnamefont {W.}~\bibnamefont
  {Streifer}},\ }\href {\doibase 10.1364/JOSA.61.000040} {\bibfield  {journal}
  {\bibinfo  {journal} {Journal of the Optical Society of America}\ }\textbf
  {\bibinfo {volume} {61}},\ \bibinfo {pages} {41} (\bibinfo {year}
  {1971})}\BibitemShut {NoStop}%
\bibitem [{\citenamefont {Deschamps}(1972)}]{Desch72}%
  \BibitemOpen
  \bibfield  {author} {\bibinfo {author} {\bibfnamefont {G.~A.}\ \bibnamefont
  {Deschamps}},\ }\href {\doibase 10.1109/PROC.1972.8850} {\bibfield  {journal}
  {\bibinfo  {journal} {Proceedings of the IEEE}\ }\textbf {\bibinfo {volume}
  {60}},\ \bibinfo {pages} {1022} (\bibinfo {year} {1972})}\BibitemShut
  {NoStop}%
\bibitem [{\citenamefont {Matichard}\ \emph {et~al.}(2015)\citenamefont
  {Matichard} \emph {et~al.}}]{aligo15_sei}%
  \BibitemOpen
  \bibfield  {author} {\bibinfo {author} {\bibfnamefont {F.}~\bibnamefont
  {Matichard}} \emph {et~al.},\ }\href {\doibase
  10.1088/0264-9381/32/18/185003} {\bibfield  {journal} {\bibinfo  {journal}
  {Classical and Quantum Gravity}\ }\textbf {\bibinfo {volume} {32}},\ \bibinfo
  {pages} {185003} (\bibinfo {year} {2015})}\BibitemShut {NoStop}%
\bibitem [{\citenamefont {Foreman}\ \emph {et~al.}(2007)\citenamefont
  {Foreman}, \citenamefont {Holman}, \citenamefont {Hudson}, \citenamefont
  {Jones},\ and\ \citenamefont {Ye}}]{ForYe07}%
  \BibitemOpen
  \bibfield  {author} {\bibinfo {author} {\bibfnamefont {S.}~\bibnamefont
  {Foreman}}, \bibinfo {author} {\bibfnamefont {K.}~\bibnamefont {Holman}},
  \bibinfo {author} {\bibfnamefont {D.}~\bibnamefont {Hudson}}, \bibinfo
  {author} {\bibfnamefont {D.}~\bibnamefont {Jones}}, \ and\ \bibinfo {author}
  {\bibfnamefont {J.}~\bibnamefont {Ye}},\ }\href {\doibase 10.1063/1.2437069}
  {\bibfield  {journal} {\bibinfo  {journal} {Review of Scientific
  Instruments}\ }\textbf {\bibinfo {volume} {78}},\ \bibinfo {pages} {021101}
  (\bibinfo {year} {2007})}\BibitemShut {NoStop}%
\bibitem [{\citenamefont {Shelby}\ \emph {et~al.}(1985)\citenamefont {Shelby},
  \citenamefont {Levenson},\ and\ \citenamefont {Bayer}}]{ShelLev85}%
  \BibitemOpen
  \bibfield  {author} {\bibinfo {author} {\bibfnamefont {R.}~\bibnamefont
  {Shelby}}, \bibinfo {author} {\bibfnamefont {M.~D.}\ \bibnamefont
  {Levenson}}, \ and\ \bibinfo {author} {\bibfnamefont {P.~W.}\ \bibnamefont
  {Bayer}},\ }\href {http://prb.aps.org/abstract/PRB/v31/i8/p5244_1} {\bibfield
   {journal} {\bibinfo  {journal} {Physical Review B}\ }\textbf {\bibinfo
  {volume} {31}},\ \bibinfo {pages} {5244} (\bibinfo {year}
  {1985})}\BibitemShut {NoStop}%
\bibitem [{\citenamefont {Wanser}(1992)}]{Wans92}%
  \BibitemOpen
  \bibfield  {author} {\bibinfo {author} {\bibfnamefont {K.}~\bibnamefont
  {Wanser}},\ }\href {\doibase 10.1049/el:19920033} {\bibfield  {journal}
  {\bibinfo  {journal} {Electronics Letters}\ }\textbf {\bibinfo {volume}
  {28}},\ \bibinfo {pages} {53} (\bibinfo {year} {1992})}\BibitemShut {NoStop}%
\bibitem [{\citenamefont {Duan}(2012)}]{Duan12}%
  \BibitemOpen
  \bibfield  {author} {\bibinfo {author} {\bibfnamefont {L.}~\bibnamefont
  {Duan}},\ }\href {\doibase 10.1103/PhysRevA.86.023817} {\bibfield  {journal}
  {\bibinfo  {journal} {Physical Review A}\ }\textbf {\bibinfo {volume} {86}},\
  \bibinfo {pages} {023817} (\bibinfo {year} {2012})}\BibitemShut {NoStop}%
\bibitem [{\citenamefont {Born}(1909)}]{Born09}%
  \BibitemOpen
  \bibfield  {author} {\bibinfo {author} {\bibfnamefont {M.}~\bibnamefont
  {Born}},\ }\href {\doibase 10.1002/andp.19093351102} {\bibfield  {journal}
  {\bibinfo  {journal} {Annalen der Physik}\ }\textbf {\bibinfo {volume}
  {335}},\ \bibinfo {pages} {1} (\bibinfo {year} {1909})}\BibitemShut {NoStop}%
\bibitem [{\citenamefont {Herglotz}(1911)}]{Herg11}%
  \BibitemOpen
  \bibfield  {author} {\bibinfo {author} {\bibfnamefont {G.}~\bibnamefont
  {Herglotz}},\ }\href {\doibase 10.1002/andp.19113411303} {\bibfield
  {journal} {\bibinfo  {journal} {Annalen der Physik}\ }\textbf {\bibinfo
  {volume} {341}},\ \bibinfo {pages} {493} (\bibinfo {year}
  {1911})}\BibitemShut {NoStop}%
\bibitem [{\citenamefont {Dalibard}\ \emph {et~al.}(1982)\citenamefont
  {Dalibard}, \citenamefont {Dupont-Roc},\ and\ \citenamefont
  {Cohen-Tannoudji}}]{DalCoh82}%
  \BibitemOpen
  \bibfield  {author} {\bibinfo {author} {\bibfnamefont {J.}~\bibnamefont
  {Dalibard}}, \bibinfo {author} {\bibfnamefont {J.}~\bibnamefont
  {Dupont-Roc}}, \ and\ \bibinfo {author} {\bibfnamefont {C.}~\bibnamefont
  {Cohen-Tannoudji}},\ }\href {\doibase 10.1051/jphys:0198200430110161700}
  {\bibfield  {journal} {\bibinfo  {journal} {Journal de Physique}\ }\textbf
  {\bibinfo {volume} {43}},\ \bibinfo {pages} {1617} (\bibinfo {year}
  {1982})}\BibitemShut {NoStop}%
\bibitem [{\citenamefont {Tolman}\ \emph {et~al.}(1931)\citenamefont {Tolman},
  \citenamefont {Ehrenfest},\ and\ \citenamefont {Podolsky}}]{Tol31}%
  \BibitemOpen
  \bibfield  {author} {\bibinfo {author} {\bibfnamefont {R.~C.}\ \bibnamefont
  {Tolman}}, \bibinfo {author} {\bibfnamefont {P.}~\bibnamefont {Ehrenfest}}, \
  and\ \bibinfo {author} {\bibfnamefont {B.}~\bibnamefont {Podolsky}},\ }\href
  {\doibase 10.1103/PhysRev.37.602} {\bibfield  {journal} {\bibinfo  {journal}
  {Physical Review}\ }\textbf {\bibinfo {volume} {37}},\ \bibinfo {pages} {602}
  (\bibinfo {year} {1931})}\BibitemShut {NoStop}%
\bibitem [{\citenamefont {Scully}(1979)}]{Scu79}%
  \BibitemOpen
  \bibfield  {author} {\bibinfo {author} {\bibfnamefont {M.~O.}\ \bibnamefont
  {Scully}},\ }\href {\doibase 10.1103/PhysRevD.19.3582} {\bibfield  {journal}
  {\bibinfo  {journal} {Physical Review D}\ }\textbf {\bibinfo {volume} {19}},\
  \bibinfo {pages} {3582} (\bibinfo {year} {1979})}\BibitemShut {NoStop}%
\bibitem [{\citenamefont {Purdy}\ \emph {et~al.}(2013)\citenamefont {Purdy},
  \citenamefont {Peterson},\ and\ \citenamefont {Regal}}]{Purdy13}%
  \BibitemOpen
  \bibfield  {author} {\bibinfo {author} {\bibfnamefont {T.~P.}\ \bibnamefont
  {Purdy}}, \bibinfo {author} {\bibfnamefont {R.~W.}\ \bibnamefont {Peterson}},
  \ and\ \bibinfo {author} {\bibfnamefont {C.~A.}\ \bibnamefont {Regal}},\
  }\href {\doibase 10.1126/science.1231282} {\bibfield  {journal} {\bibinfo
  {journal} {Science}\ }\textbf {\bibinfo {volume} {339}},\ \bibinfo {pages}
  {801} (\bibinfo {year} {2013})}\BibitemShut {NoStop}%
\bibitem [{\citenamefont {Wilson}\ \emph {et~al.}(2015)\citenamefont {Wilson},
  \citenamefont {Sudhir}, \citenamefont {Piro}, \citenamefont {Schilling},
  \citenamefont {Ghadimi},\ and\ \citenamefont {Kippenberg}}]{WilSud15}%
  \BibitemOpen
  \bibfield  {author} {\bibinfo {author} {\bibfnamefont {D.~J.}\ \bibnamefont
  {Wilson}}, \bibinfo {author} {\bibfnamefont {V.}~\bibnamefont {Sudhir}},
  \bibinfo {author} {\bibfnamefont {N.}~\bibnamefont {Piro}}, \bibinfo {author}
  {\bibfnamefont {R.}~\bibnamefont {Schilling}}, \bibinfo {author}
  {\bibfnamefont {A.}~\bibnamefont {Ghadimi}}, \ and\ \bibinfo {author}
  {\bibfnamefont {T.~J.}\ \bibnamefont {Kippenberg}},\ }\href {\doibase
  10.1038/nature14672} {\bibfield  {journal} {\bibinfo  {journal} {Nature}\
  }\textbf {\bibinfo {volume} {524}},\ \bibinfo {pages} {325} (\bibinfo {year}
  {2015})}\BibitemShut {NoStop}%
\end{thebibliography}%

\end{document}